**Intravital imaging and cavitation monitoring of antivascular ultrasound in tumor microvasculature**
*Xiaoxiao Zhao*[*,1,2], Carly Pellow*[*,2], David E. Goertz[1,2]*

[1]Department of Medical Biophysics, University of Toronto, Canada; [2]Sunnybrook Research Institute, Toronto, Canada
*These authors contributed equally to this work.

Corresponding author: Xiaoxiao Zhao, E-mail: xxiao.zhao@mail.utoronto.ca, and David Goertz, E-mail: goertz@sri.utoronto.ca, 2075 Bayview Ave, Toronto ON, M3C 3Z6

**ABSTRACT**
**Rationale:** Focused ultrasound-stimulated microbubbles have been shown to be capable of inducing blood flow shutdown and necrosis in a range of tissue types in an approach termed antivascular ultrasound or mechanical ablation. In oncology, this approach has demonstrated tumor growth inhibition, and profound synergistic antitumor effects when combined with traditional platforms of chemo-, radiation- and immune-therapies. However, the exposure schemes employed have been broad and underlying mechanisms remain unclear with fundamental questions about exposures, vessel types and sizes involved, and the nature of bubble behaviors and their acoustic emissions resulting in vascular damage – impeding the establishment of standard protocols.
**Methods:** Here, ultrasound transmitters and receivers are integrated into a murine dorsal window chamber tumor model for intravital microscopy studies capable of real-time visual and acoustic monitoring during antivascular ultrasound. Vessel type (normal and tumor-affected), caliber, and viability are assessed under higher pressure conditions (1, 2, and 3 MPa), and cavitation signatures are linked to the biological effects.
**Results:** Vascular events occurred preferentially in tumor-affected vessels with greater incidence in smaller vessels and with more severity as a function of increasing pressure. Vascular blood flow shutdown was found to be due to a combination of focal disruption events and network-related flow changes. Acoustic emissions displayed elevated broadband noise and distinct sub- and ultra-harmonics and their associated third-order peaks with increasing pressure.
**Conclusions:** The observed vascular events taken collectively with identified cavitation signatures provide an improved mechanistic understanding of antivascular ultrasound at the microscale, with implications for establishing a specific treatment protocol and control platform.

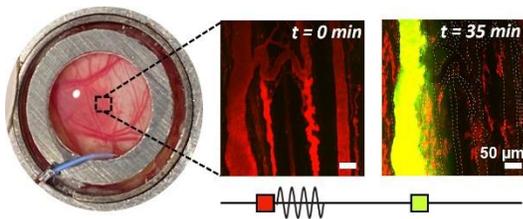

*Graphical abstract: Antivascular ultrasound is monitored in real-time with simultaneous acoustic monitoring and multiphoton microscopy in a murine dorsal window chamber tumor model.*

**INTRODUCTION**
Therapeutic focused ultrasound (FUS) is rapidly emerging as a tool to noninvasively induce controlled and targeted biological effects, with potential to impact a wide range of clinical applications, including the treatment of tumors. While a range of ultrasound methods can be employed, a prominent approach involves the use of exogenous microbubbles (MBs) in the blood stream, which are focally stimulated by FUS. The premise of MB-mediated therapy is that MBs entering the localized region of the FUS beam undergo pressure-dependent oscillations and interact with tissue at time-averaged acoustic power levels that are over two orders of magnitude lower than the threshold without them [1, 2], and at more predictable pressures [3]. Compelling studies have highlighted the ability to elicit enhanced microvascular permeability [4, 5], inflammation [6], apoptosis [7], necrosis [8], thrombolysis [9, 10], thrombogenesis [11], angiogenesis [12], vessel rupture [13], and vascular shutdown [14, 15].

These effects are dependent on exposure conditions and have been linked to MB behaviors and their interactions with vessel walls [16]. The most developed application of MB-mediated FUS is drug delivery, where pressures are typically limited to 100s of kPa and MB-induced microstreaming and mild endothelial cell deformation [4] gives rise to enhanced microvascular permeability and extravasation of circulating therapeutic agents. Importantly, these behaviors broadly correspond to emitted acoustic signatures, providing a means to enable non-invasive monitoring and facilitate treatment control [17, 18, 19] and clinical translation. When the objective is to deliver drugs while preserving the target tissue, it is also desirable to avoid inducing inflammation or damage (i.e. erythrocyte extravasation) that can occur at higher MB oscillation amplitudes [11, 20]. The latter effects have been associated with broadband (inertial cavitation) or threshold-dependent sub- or ultra-harmonic emissions. These principles have been leveraged to transiently disrupt the blood-brain barrier to facilitate drug delivery to the brain, which entered clinical trials with a hemispherical transcranial array FUS system in 2015 [21, 22, 23], controlled by monitoring stable MB cavitation including the subharmonic [21], ultraharmonics [19], or a combination of harmonic and broadband emission [24]. Initial clinical trials have also been conducted with the intention of enhancing drug delivery to pancreatic tumors [25, 26], malignant neoplasms of the digestive system [27], and liver metastases from beast and colorectal cancers [28, 29], though notably these were carried out with conventional diagnostic ultrasound systems in the absence of cavitation monitoring and control.

Another approach for FUS MB-mediated therapy is to induce microvascular shutdown of blood flow and thereby tissue necrosis, a method that has been referred to as mechanical ablation, nonthermal ablation, or antivascular ultrasound (AVUS) [14]. This approach is less studied than drug delivery and, as discussed below, effects typically occur at higher pressures (on the order of a few MPa, though tissue type, cavitation agent employed, and other pulse parameters are also determining factors) and may be linked to the presence of inertial cavitation [15, 30, 31]. AVUS has been demonstrated preclinically in a range of tissue types [32, 33, 31, 34, 35, 36, 37] and has potential for applications where tissue destruction is desirable, such as in selected brain regions or tumors, the latter being of interest in the present study. While AVUS has demonstrated tumor growth inhibition on its own, the most profound antitumor effects are elicited when combining with conventional approaches of chemotherapy [38, 15, 30], antiangiogenic therapy [39, 40], immunotherapy [41, 42], and radiotherapy [43, 44, 45, 46, 47, 48].

When considered from a drug delivery perspective, the use of AVUS may at first appear counterintuitive in that disrupting microvessels would reduce drug access to the tumor. However, it has been proposed that AVUS may act as a physical analog to small molecule vascular disruption agents [49, 50] which have a long history of being combined with conventional therapies – albeit with a significant side effect profile [51, 15]. A biological rationale for combining vascular disrupting agents with chemotherapy is that vascular shutdown preferentially induces necrosis in the tumor core where more fragile angiogenic neovasculature is present, while chemotherapeutics affect the well-perfused tumor rims where more robust normal and co-opted host vessels are present [52]. This pattern has been shown to hold with AVUS [30, 15, 41, 53, 54]. With antiangiogenic strategies, the rationale is that antiangiogenic treatments block vascular rebound [55], which is consistent with results reported for AVUS with metronomic chemotherapy [30] and Sorafenib [56]. AVUS has also been combined with anti-PD-1 checkpoint blockade, where the combination treatments significantly inhibited tumor growth and enhanced survival relative to monotherapies and may be linked to initiating an adaptive immune response [41]. Finally, there is a substantial body of work demonstrating that AVUS can profoundly enhance the antitumor effects of radiation therapy [43, 44, 45, 46, 47, 48, 57]. Here it is posited that MB stimulation mechanically perturbs endothelial cell membranes activating ceramide-based biomechanical pathways that enhance endothelial cell radiosensitivity, and radiation induces secondary cell death by ceramide-related endothelial cell apoptosis leading to vascular shutdown. The combination of AVUS and radiation therapy has been the first to emerge into clinical trials for head and neck cancer, and chest-wall and locally advanced breast cancer as of 2020 [58].

While the results to date suggest considerable potential for AVUS, relative to FUS MB-mediated drug delivery it is less well understood in terms of mechanisms, appropriate exposure conditions, and it lacks implementation of validated cavitation-based control methods. An overview of preclinical experiments studying AVUS in various tumor and tissue types is provided in *Table S1*. FUS exposures have utilized a range of frequencies (0.24 – 5 MHz), peak negative pressures from 0.18 – 10 MPa, modes of continuous wave and pulsed waves with a burst duration of 0.0015 – 100 ms, and pulse repetition frequencies (PRF) of 0.01 – 3 kHz; although within this parameter space only sparse exploration has been conducted. Exposure approaches have included continuous wave unfocused physiotherapy systems, diagnostic ultrasound machines, and single-element fixed or scanned focused transducers for pulsed wave implementations. A variety of commercial and in-house MB types have been employed, with doses ranging from 1.5-3400x clinical diagnostic ultrasound dosing. It is possible that the range of parameters yield different mechanisms of vascular shutdown: For example, some studies indicate MB-induced endothelial damage followed by platelet aggregation and vessel thrombosis [59, 60, 35], while others report blood flow shutdown in the absence of hemorrhage or endothelial cell death [32, 61]. Similarly, studies have monitored vascular shutdown on different timescales, with immediate investigations of perfusion [42, 62, 63, 30, 15, 41] or later surrogate indications of endothelial apoptosis [64, 65, 43]. In addition, few studies have monitored cavitation during AVUS therapy [41, 15, 30, 66, 67] to formally link cavitation signatures to vascular shutdown.

Therefore, fundamental questions remain about the vessel sizes and types involved, timescale of effects, the nature of MB behaviors required to induce damage, and their associated acoustic signatures. In the present study we integrate ultrasound transmitters and receivers into a murine dorsal window chamber model for intravital microscopy studies capable of real-time visual and acoustic monitoring during and following FUS-induced MB cavitation. Vessel type (normal and tumor), caliber, and viability are assessed under therapeutic ultrasound conditions (1, 2, and 3 MPa to investigate pressure-dependence of effects), and cavitation signatures are linked to the biological effects. A 1 MHz transmit frequency millisecond scale pulsing regimen is employed, with a view towards the study information being compatible with array-based therapy systems currently under development. An improved mechanistic understanding of AVUS at the microscale is elucidated, with implications for establishing an AVUS-specific treatment protocol and control platform.

## MATERIALS AND METHODS

**Tumor cell line and animal preparation.** Green fluorescent protein tagged human FaDu squamous cell carcinoma cells (FaDu-GFP; AntiCancer Inc.) were cultured at 37 °C in a mixture of 5% $CO_2$ and 95% air. Cells were propagated in RPMI medium 1640 with L-glutamine (MultiCell Technologies Inc.), supplemented with 10% FBS, 100 U/mL penicillin, and 100 µg/mL streptomycin, and were trypsinized and harvested prior to reaching confluency. Six to eight-week old BALB/c nude or athymic nude mice (Charles River Laboratories) underwent surgery for dorsal skinfold window chamber and tumor cell implantation (as in [68, 69, 70]): A titanium frame (APJ Trading Co. Inc.) was surgically implanted to hold the dorsal skinfold within a transparent window. The upper layer of the dorsal skinfold was removed, and $2x10^6$ FaDu-GFP tumor cells suspended in 30 µL of media were injected with a 30 G needle into the fascia in the centre of the window. A 12 mm diameter glass coverslip was then placed over the exposed skinfold, and mice were monitored until imaging studies 7-9 days later. All animal procedures were approved and conducted in compliance with the Animal Care Committee guidelines at Sunnybrook Research Institute, Canada.

For imaging, mice were prepared as in [68, 69, 70] (*Figure 1A-B*). Briefly, mice were anesthetized with isofluorane with a 2:3 ratio of oxygen and medical air carrier gases. Tail veins were cannulated with a 27 G catheter, and mice were placed on a heating pad on a removable microscope stage with rectal thermistor feedback to maintain a core body temperature of 37 °C (TC-1000; CWE Inc.). The window chamber coverslip was removed, the skinfold was wet with degassed saline, and a 12 mm diameter, 150 µm thick glass coverslip with an in-house zirconate titanate PZT-4 ring transducer affixed to the top surface with cyanoacrylate adhesive was placed over the exposed skin and held with a retaining ring. The underside of the window chamber was coupled by ultrasound gel to a degassed water bath reservoir heated with a circulating water heater (T/Pump model TP-500; Gaymar Industries Inc.) to maintain the window chamber temperature at 37 °C. On the bottom of the water reservoir was an in-house polyvinylidene fluoride (PVDF) receive transducer used for passive cavitation detection. The removable microscope stage was then transferred to the multiphoton microscope for the study.

**Multiphoton microscopy settings.** The dorsal window chamber was placed under a water immersion objective lens (25x 1.05 NA; XLPLN25xWMP2, Olympus) with a field-of-view (FOV) of 500 µm x 500 µm. Laser scanning was performed using a multiphoton microscope (FV1000MPE; Olympus), exciting the sample at 900 nm with a mode-locked Titanium Sapphire tunable laser (690-1040 nm; MaiTai, Spectra-Physics). Fluorescent emissions following bandpass filtering were collected by a photomultiplier tube following bandpass filtering of 575-645 nm for tetramethylrhodamine (TMR dextran, 2 MDa; Invitrogen), and 495-540 nm for fluorescein isothiocyanate (FITC dextran, 2 MDa; Invitrogen) and FaDu-GFP cells.

**Ultrasound parameters.** In-house PZT-4 ring transducers were pressure calibrated with a fiber-optic hydrophone (Precision Acoustics) prior to experiments (*Figure S1*): The ring transducers (10 mm outer diameter, 8.5 mm inner diameter, 1.1 mm thickness) were affixed onto a 12 mm diameter, 150 µm thick glass coverslip with cyanoacrylate adhesive [71] (*Figure 1C*) and held in a window chamber by an internal retaining ring on the surface of a degassed, deionized water bath. A droplet of degassed, deionized water was placed in the inner part of the ring and coupled with an objective lens. A frequency sweep from 1-1.1 MHz in increments of 0.05 MHz was performed in thickness mode to achieve a minimum pressure change over 1-200 µm depth below the coverslip, corresponding to the maximum imaging depth (with a pressure change tolerance of <15%). On average, the selected driving frequency was 1.05 ± 0.05 MHz, and the maximum peak negative pressure was at a depth of 50 µm below the coverslip.

For experiment sonication, the ring transducer was air-backed, with a droplet of degassed water in the inner part of the cylinder for compatibility with the water-immersion lens. Two arbitrary waveform generators (AFG3022B; Tektronix) were used to drive the ring transducer. One of these was used to control pulsing parameters, while the second one was used to trigger pulsing for a set number of repetitions: FUS exposures were performed at the ring transducer's fundamental frequency with a pulse length of 5 ms, pulse repetition period of 10 s to enable inter-pulse reperfusion, for a 3 min duration at 1, 2, or 3 MPa peak negative pressure. The pressures employed are within range of previous AVUS studies and were refined based on pilot work conducted as part of this study. These pulses were generated by the arbitrary waveform generators, attenuated by 20 dB (HAT-20+, DC-2GHz, MiniCircuits), amplified with a 55 dB RF power amplifier (model A150, 0.3-35MHz, 150W, Electronics & Innovation), and transmitted through the ring transducer. Passive cavitation detection was performed with an in-house broadband PVDF receiver (4 mm diameter active element, *Figure 1D*) approximately 1.5 cm below the coverslip, centred at 10 MHz. Received signals were passed through a preamplifier and recorded for the entirety of the ultrasound pulses and were digitized at 125 MHz with a 14-bit PC-based oscilloscope (PicoScope 6402C, Pico Technology Ltd.).

**Experimental procedure.** The objective lens and ring transducer were coaligned such that imaging captured the sonication focus in the centre of the FOV. Following the procedure outlined in *Figure 1E*, TMR-dextran was first injected (6 mL/kg at a concentration of 5 mg/mL *via* tail vein catheter) to visualize the vasculature and adjust the multiphoton microscope excitation laser power and detector channel HV, gain and offset settings for imaging over depth. Imaging was completed from the surface of the dorsal skinfold to a depth of up to 250 µm to maintain a high signal-to-noise ratio for the microscope and remain close to the acoustic focus (for up to a 15% change in pressure). A baseline XYZ volume scan was acquired to create a 3D vascular map prior to treatment. Volume scans were acquired with 512x512 pixels (509.12 µm x 509.12 µm, resolution of 0.9944 µm/pixel) to a depth of up to 250 µm in 1 µm increments at 2 µs/pixel.

A tissue plane of depth between 30-75 µm was then selected such that vessels of various sizes could be visualized with good SNR and such that the FOV corresponded to approximately the maximum pressure. In the chosen plane, a baseline XYT time scan and baseline cavitation data were acquired with the same pulse sequence as would be utilized later for therapy. Single plane XYT time scans were acquired with the same lateral and temporal resolution as the volume scans, with a total imaging duration of 5 min to capture the duration of the bubble injection and sonication. Definity® MBs (Lantheus Medical Imaging Inc.) were then prepared as follows: MB vials were activated at room temperature in a VialMix (Lantheus Medical Imaging Inc.) for 45 s and then allowed to passively cool to room temperature. The MBs were then gently resuspended and allowed to decant for 30 s prior to extracting 50 µL for subsequent dilution in saline to a dosage of 60 µL/kg for injection through the tail vein catheter. This MB dose is equivalent to 3x the clinical imaging dose and is at the lower end of the range utilized in preclinical AVUS experiments. Another time scan was then initiated at the same plane as the baseline scan, and ultrasound exposures were transmitted 10 s after MB injection.

At 5 min and 30 min post-sonication, volume scans were acquired to assess vascular damage and dye leakage. FITC-dextran was then injected (6 mL/kg at a concentration of 5 mg/mL *via* tail vein catheter) and volume scans were acquired to distinguish shutdown vessels at 35 min post-treatment and to assess potential recovery at 60 min post-treatment.

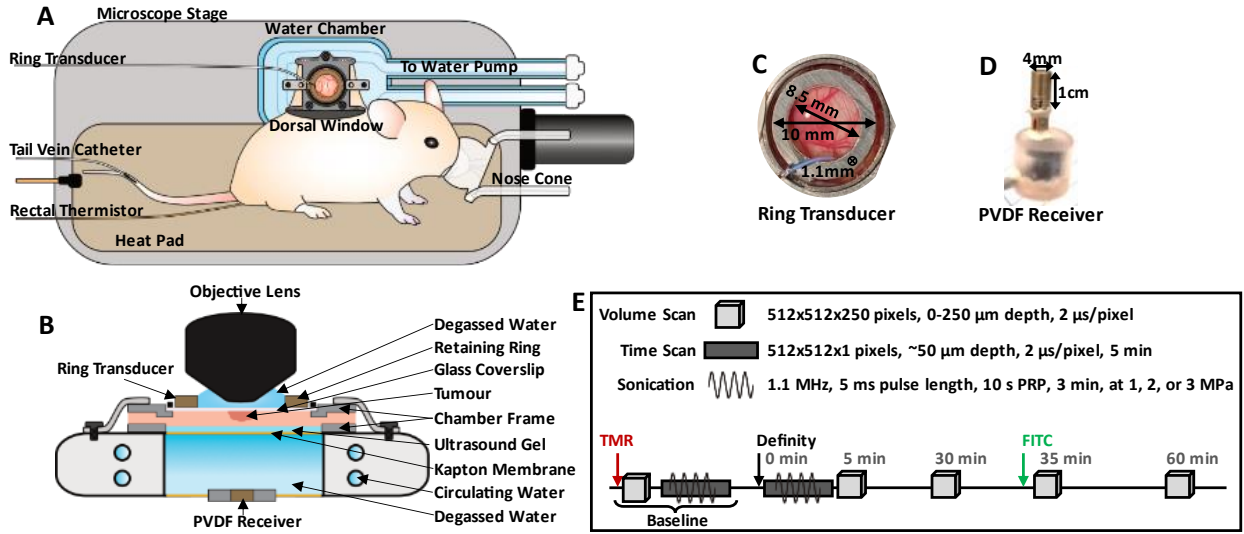

*Figure 1. Overview schematic of the integrated acoustic and microscopy dorsal window chamber setup.* (**A**) Top-view sketch of a mouse with a dorsal window chamber prepared for experiments on a removable microscope stage. (**B**) Side-view of the main experimental region; the dorsal chamber and objective lens, with a ring transducer and PVDF receiver integrated for acoustic and visual tracking capabilities. (**C**) Photograph of the ring transducer held in the dorsal window chamber, with dimensions. (**D**) Photograph of the PVDF receiver used to detect acoustic scattering on the opposite side of the dorsal chamber as the ring transducer, with dimensions. (**E**) Scheme utilized to monitor spatiotemporal leakage upon FUS stimulation of circulating Definity MBs, determine vascular shutdown, and monitor acoustic emissions at 1, 2, or 3 MPa peak negative pressure in mice.

**Multiphoton microscopy data analysis.** *Spatiotemporal extravasation:* Multiphoton fluorescence images were first assessed in MATLAB (MathWorks) for dye extravasation. The 575-645 nm channel (corresponding to the TMR dextran) was corrected for GFP bleed-through, median filtered in 3-dimensions, and contrast enhanced *via* contrast-limited adaptive histogram equalization. A binarized 3D vessel mask was then created through iterative thresholding to segment the intra- and extra-vascular compartments in the baseline XYZ volume stack (prior to MB injection and sonication). Average dye concentration changes in the intra- and extra-vascular compartments were then examined by applying the 3D mask to subsequent times (5, 30 min post-FUS), normalized to the compartmental volume and baseline fluorescence with the assumption that fluorescence is proportional to concentration. The area under the concentration-time curve (AUC) was then calculated by trapezoidal integration as a function of time to yield a measure of bioavailability. Significance between pressure groups and compartments was evaluated with one-tail, two-sample, equal variance t-tests. Spatiotemporal release was then assessed as follows. A Euclidean distance transform of the segmented 3D vascular map was performed to create a distance map from each extravascular pixel to the nearest vascular structure, and applied to the extravascular compartment at baseline (0 min), and at 5 and 30 min post-FUS. Boundary effects were removed by truncating the volume by 20 pixels on all sides. Extravascular compartment fluorescence was normalized with respect to the compartment volume at each distance away from the nearest vessel, and to baseline fluorescence.

*Vessel measurements, events, and shutdown assessment:* Vessels were counted and sized in FIJI (ImageJ). Vessels were segmented by nodes such that any bifurcation divided the original vessel into multiple at the node. Each vessel was given a tracking identification and its diameter was measured by taking three measurements of the maximum length of the signal profile perpendicular to the local tangent along the length of the vessel and averaged. Vessels were tracked across multiple time points to assess changes upon sonication: Vasospasm was identified when deformations in vessel length were visible. Vasoconstriction and vasodilation were noted when vessel diameter constricted or dilated by at least 10%. Flow reversal was classified by tracking a reversal in the bulk motion of the fluorescent dextran or dark voids of red blood cells through vessels. Vascular occlusions or clots were identified when red blood cells were observed to form large clots that moved slowly or became trapped in vessels. Focal disruption or leakage was classified upon visualization of distinct fluorescent dextran extravasation, visualized first during the sonication in the XYT time scan or in the 5 min XYZ volume scan, and becoming more diffuse over subsequent imaging timepoints. Following sonication, the images upon FITC-dextran injection (35, 60 min post-FUS) were utilized to assess whether vascular shutdown occurred. Injection of this second dye was necessary to enable clearer visualization of shutdown, as diffuse TMR dye from leakage events as well as slower flow in some vessels frequently precluded the reliable assessment of shutdown with one dye alone. Vessels were classified as viable if the FITC-dextran flowed through them (i.e. they appeared yellow or green) and shutdown if they remained red or appeared dark at 35 min. Shutdown was then categorized as transient if the vessel blood flow recovered, or sustained if the vessel remained red or dark at 60 min. Vascular events were normalized to the total number of vessels visualized at each pressure. Significance of shutdown levels between groups was evaluated with either paired t-tests, two-sample equal variance t-tests, or two-sample unequal variance t-tests as appropriate (depending on sample sizes and variance as evaluated with an f-test).

**Cavitation data analysis.** Acoustic data was post-processed in MATLAB (MathWorks). Received signals (baseline acquired prior to agent injection, and bubble cavitation acquired upon MB injection and sonication) were digitally filtered by a $5^{th}$ order bandpass Butterworth filter (0.3 MHz high-pass and 10 MHz low-pass) and were then processed in two different ways. The first method assessed the entire received signal burst-by-burst by multiplying by a 5 ms rectangular window centred on the received signal. The second method assessed the cavitation response over time throughout the first burst by multiplying by a moving 100 µs Hanning window. The moving window began at the beginning of the received signal and moved by 50 µs increments (50% overlap) to reach the end of the 5 ms received signal. For both methods, windowed received signals were zero-padded to a frequency resolution of 200 Hz per division prior to computing the Fourier transform.

Example spectra (*Figure 5*) are composed of the averaged power spectra of the first two bursts within the first millisecond window for both baseline and bubble cavitation signals, both normalized to the mean power between 3.1 and 3.4 times the fundamental frequency of the first burst in the baseline signal for each mouse. Quantified cavitation data (*Figure 6*, *Figure S3*) was calculated by integrating the power spectra (obtained with the 5 ms window on the first burst) over the following frequency bands of interest; inertial cavitation IC (mean over $(1.9-1.95)f_1$ and $(2.05-2.1)f_1$), subharmonic $½f_1$ (mean of 4 kHz bandwidth around the peak point value), $1/3+2/3$ subharmonic (mean of 4 kHz bandwidth around $(1/3)f_1$ and $(2/3)f_1$), relative subharmonic (subtraction of the surrounding broadband mean power at $(0.43-0.46)f_1$ and $(0.54-0.57)f_1$ from the subharmonic), ultraharmonic $3/2f_1$ (mean of 4 kHz bandwidth around the peak value), $1/3+2/3$ ultraharmonic (mean of 4 kHz bandwidth around $(4/3)f_1$ and $(5/3)f_1$), and the relative ultraharmonic (subtraction of the surrounding broadband mean power at $(1.43-1.46)f_1$ and $(1.54-1.57)f_1$ from the ultraharmonic). As discussed later, the use of these specific third-order sub- and ultra-harmonic frequencies is atypical and is based on the spectral observations that emerged. Quantified cavitation data involves the subtraction of baseline signals from bubble cavitation signals, except for the 'relative' bands. Statistical significance on quantified cavitation data above was calculated by performing right-tail one-sided t-tests. Quantified cavitation data was further provided as a function of burst number (with the first assessment method above of burst-by-burst 5 ms rectangular windowing), and within the first burst (with the second method above using a moving 100 µs Hanning window) to indicate persistence (*Figure S4*). Inertial cavitation dose was calculated by summing the quantified cavitation data in this band across all 18 bursts of the treatment duration for each mouse (for the bubble cavitation data). The average and standard deviation of the inertial cavitation dose was then calculated for each pressure group. Statistical significance between pressure groups was calculated by performing one-tail, two-sample, equal variance t-tests on the integrated power.

**RESULTS**
**Effects induced during therapeutic ultrasound exposure.** A range of vascular events were visualized upon ultrasound exposure (1.05 MHz, 5 ms pulse length, 10 s pulse repetition period, 3 min duration at 1, 2, or 3 MPa peak negative pressure) during XYT time scans as well as immediately following sonication through XYZ volume scans at 5 min, depicted in *Figure 2*. Video examples of events occurring during XYT time scans can be found in the ***Supplementary Information***. Events were typically more prominent as a function of increasing pressure (*Figure 2A*).

Rapid-onset events captured within the first 30 s of sonication during the XYT time scans included vasospasm, vasoconstriction, and vasodilation (*Figure 2B-D*), each of which involved smaller vessels with greater severity. In many cases where sudden (between frame) focal disruption occurred, this was accompanied by substantial and diffuse tissue motion, where tissue typically did not fully relax to its prior state for a number of frames. Such motion was also evident for events that appeared to occur outside of the optical focal plane (See *Videos S1-8*). Frequently following these events, focal disruptions resulting in localized dye leakage could be observed, typically occurring within the first minute of sonication near bifurcations (*Figure 2E*). Formation of clots or occlusions could be visualized close to points of focal disruption later during sonication but was not a common occurrence (*Figure 2F*). Throughout the exposure duration though, flow directionality reversal could be observed (*Figure 2G*), sometimes preceding focal disruptions (though with the limitation that only a single plane could be visualized at this temporal resolution during sonication) and often following them. Flow reversals occurred not only in the vessels experiencing focal disruptions but in up- and down-stream connected vessels in the network. Vessels experiencing focal disruptions were also prone to exhibiting blood flow shutdown (*Figure 2H*). This event could be observed towards the end of sonication with vessels appearing dark or with stagnant erythrocytes, and could be more easily identified with the addition of a second dye (FITC dextran) at 30 min following sonication. Notably, many vessels up- and down-stream in the network from a focal disruption experienced flow reversal followed by flow shutdown, though sustained events of shutdown (60 min) were more likely to be close to the original disruption location. The most prominent events of focal disruption and shutdown will be further analyzed in the following section.

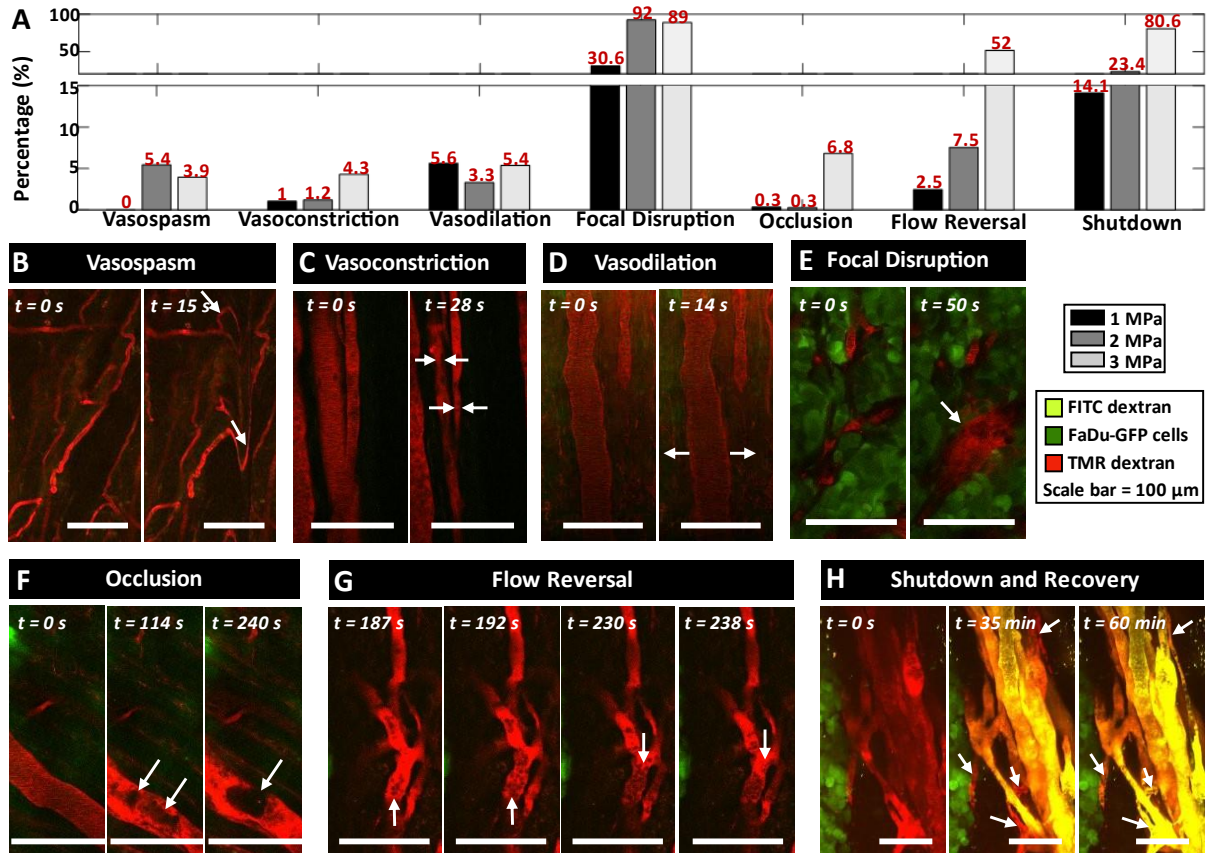

*Figure 2. Microvascular events induced by vascular disruption therapy visualized with intravital multiphoton microscopy. Fluorescent FaDu-GFP tumor cells are visualized in green, injected 2 MDa TMR-dextran is red, and injected 2 MDa FITC-dextran is yellow. The scale bars represent 100 µm. White arrows indicate the events. (A) Relative occurrences of various vascular events visualized at 1, 2 and 3 MPa. Events are normalized to the total number of vessels visualized at each pressure. (B) Vasospasm seen as twisting and displacement of vessels in the first 15 s of sonication (XYT time scan). (C) Constriction and (D) dilation of vessels, often seen within the first 30 s of sonication (XYT time scan). (E) Throughout the duration of sonication vascular disruptions were observed including focal and more diffuse leakage, frequently following the above-mentioned events (or preceding, in the case of vascular shutdown) when they occurred (typically within the first 1 min of sonication; XYT time scan). (F) Formation of clots occluding a vessel, occurring slowly and towards the end of sonication (XYT time scan). (G) Red blood cells visualized flowing towards and then away from a node in an apparent flow reversal (XYT time scan). (H) Vascular blood flow was observed to stop for at least 35 min after sonication, and in some cases recover by 60 min (visualized through XYZ volume scans at 0, 35, and 60 min following sonication).*

**Vascular leakage and blood flow shutdown following antivascular ultrasound.**
The most prominent vascular event to occur during ultrasound exposure was focal disruption, also referred to here as leakage, with frequent occurrences of subsequent shutdown. Leakage was first visualized as being focal during sonication throughout the XYT time scan, consistent with association with a local bubble event. It was then found to plateau and become more diffuse through subsequent XYZ volume scans at 5 and then 30 min following sonication. Focal disruption events were found to exhibit increasing extravasation distance and quantity with rising pressure (*Figure 3A-B*). Of note is the fact that vessels experiencing focal disruptions also frequently exhibited blood flow shutdown (*Figure 3C*), evident at 35 min following sonication. Specifically, of the total vessels that experienced leakage, 16% also exhibited shutdown at 1 MPa; 37% at 2 MPa; and 79% at 3 MPa. Both leakage and shutdown preferentially occurred in smaller vessels (<20 µm diameter), and progressively affected larger and more vessels with increasing pressure (*Figure 3C-D*). It is also notable that with increasing pressure, up- and down-stream vessels networked with the focally disrupted vessel progressively also sustained flow shutdown (resulting in a marginal increase in total leakage events, but a substantial increase in shutdown vessels: Of the vessels experiencing shutdown, 22% also had a leakage event at 1 MPa; 70% at 2 MPa; 52% at 3 MPa). It was also observed that not all vessels with leakage events also exhibited shutdown and vice versa, but there was substantial overlap with increasing pressure (*Figure 3D, Figure S2*). The mean vessel sizes shutdown (with or without leakage) were also similar to those with leakage and indeed the overall examined vessel mean (*Table S2*).

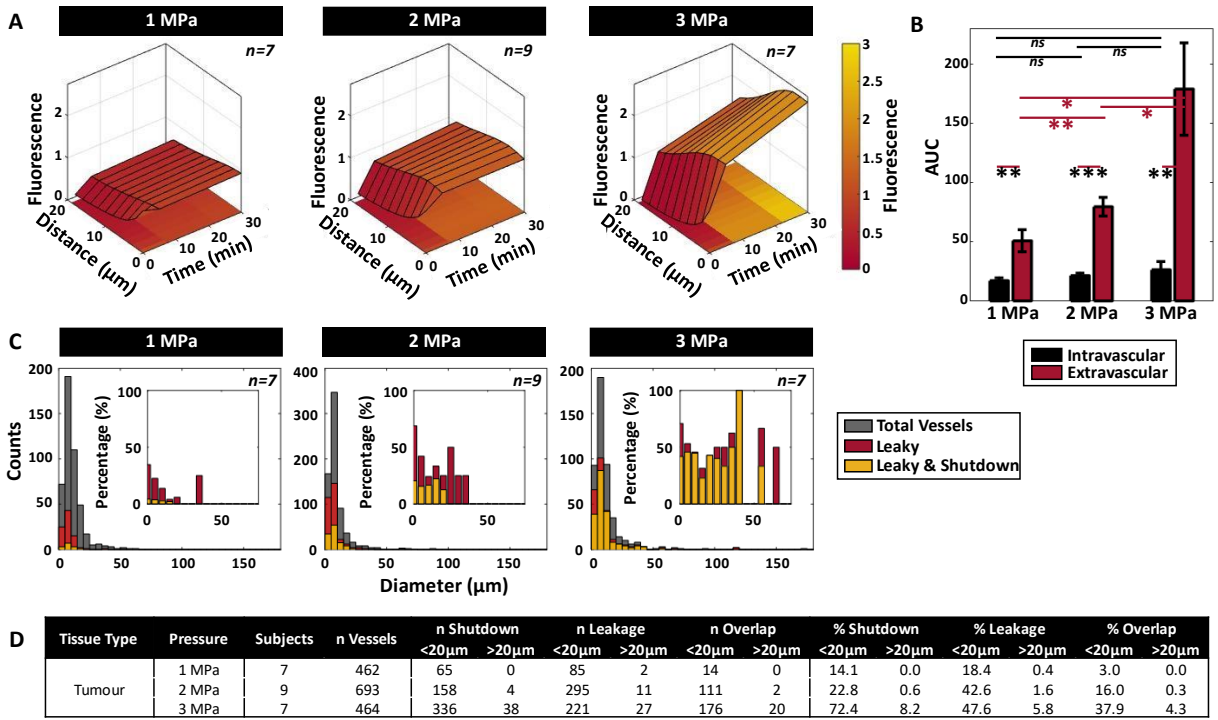

*Figure 3. Spatiotemporal extravasation, bioavailability, and enhanced vascular permeability and shutdown following ultrasound exposure as a function of pressure. (A) Mean extravasation distance surface plots from the nearest vessel as a function of time in tumor-bearing mice treated at 1 (n=7), 2 (n=9) and 3 (n=7) MPa from XYZ volumetric scans. (B) Area under the curve (AUC) measurements (mean and standard deviation), demonstrating increasing extravascular bioavailability of injected TMR-dextran (as a drug surrogate) after ultrasound exposure. \* p<0.05, \*\* p<0.01, \*\*\* p<0.001, n.s. = not significant. (C) Histograms of absolute counts of all visualized vessels overlayed by vessels exhibiting leakiness as well as both leakiness and flow shutdown at 1, 2, and 3 MPa in mice with tumors. Histograms of percentage of leaky as well as leaky and shutdown vessels normalized to total visualized vessels at 1, 2, and 3 MPa are inset. (D) Summary of shutdown and leakage occurrences, as well as the overlap of these events in small (<20 µm) and large (>20 µm) diameter vessels. Percentages are provided normalized to the total number of vessels studied at each pressure.*

*Figure 4A* displays histograms of vessels with flow shutdown induced by ultrasound, overlayed with a subset of vessels that recovered flow within 1 h after treatment. Shutdown occurred preferentially in smaller vessels, affecting larger and more vessels with increasing pressure (the largest being 71 µm diameter). Recovery occurred preferentially at lower pressures and in smaller vessels (*Table S3*). For comparison, normal mice (sham surgeries) were utilized as controls. The same trends applied to the normal mice, but with overall lower incidence suggesting a self-targeting effect to tumors. Consideration of all vessels (i.e. without binning) demonstrates further the propensity of vascular shutdown in tumors compared to normal tissue, as well as the tendency of shutdown vessels to recover at lower pressures and preferentially in normal tissue (*Figure 4B-C*, *Table S3*).

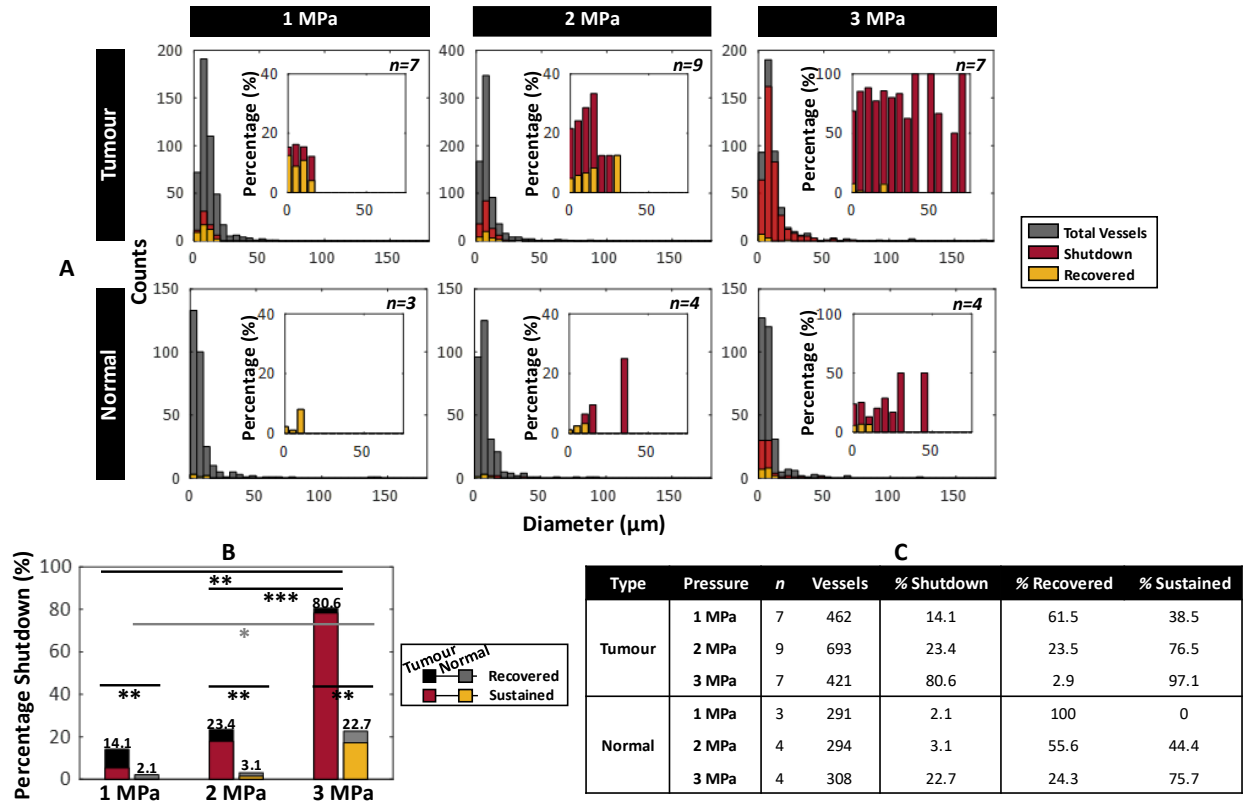

*Figure 4. Blood flow shutdown induced by ultrasound in tumor-bearing and normal mice. (A) Histograms of absolute counts of all visualized vessels overlayed by vessels exhibiting shutdown and recovery at 1, 2, and 3 MPa in tumor-bearing and normal mice. Histograms of percentage of shutdown vessels overlayed with recovered vessels normalized to total visualized vessels at 1, 2, and 3 MPa are inset. (B) Total percentage shutdown in tumor and normal tissue at 1, 2, and 3 MPa. Stacked colours indicate the relative breakdown of shutdown vessels that recover (black in tumors, grey in normal tissue) and exhibit sustained shutdown (red in tumors, yellow in normal tissue). For tumor tissue, increasing pressure results in significant increases in vascular shutdown, whereas in normal tissue significance is not reached until 3 MPa. $p<0.05$ = *, $p<0.01$ = **, $p<0.001$ = ***. (C) Summary of total number of subjects and vessels studied, along with relative percentages of vessels exhibiting shutdown. Of the shutdown vessels, relative percentages of recovered and sustained shutdown vessels are provided.*

**Cavitation signatures linked to antivascular ultrasound.**

Acoustic emissions were monitored during treatment to link cavitation signatures to vascular shutdown (*Figure 5*). At 1 MPa, peaks at the fundamental and second harmonic were detected along with a degree of inertial cavitation. At 2 MPa, sub- and ultra-harmonic peaks began to emerge along with peaks at 1/3 and 2/3 of the sub- and ultra-harmonics respectively and broadband noise. At 3 MPa, sub- and ultra-harmonic peaks as well as their 1/3 and 2/3 peak counterparts became more distinctive above the broadband noise. Notably, the microscopy data indicated a higher degree of vascular shutdown with elevated pressure, signifying the possibility of utilizing broadband noise as well as the sub- and ultra-harmonics (and their 1/3, 2/3 elevated counterparts) as indicators for antivascular ultrasound therapy.

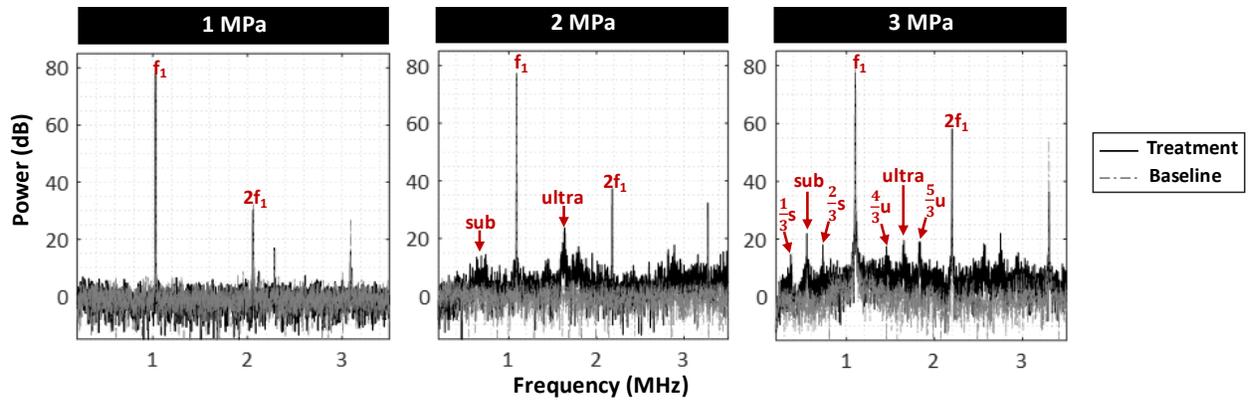

*Figure 5. Sample cavitation spectra ultrasound exposure.* The fundamental frequency ($f_1$) and second harmonic ($2f_1$) are present at 1, 2, and 3 MPa. As pressure increases, broadband noise as well as distinct sub- and ultra-harmonics emerge, along with sharp peaks at 1/3 and 2/3 of the sub- and ultra-harmonics.

Quantifying cavitation data in specific frequency bands across mice and pressure groups in tumor vasculature (*Figure 6*), it was found that harmonic peaks (subharmonic, ultraharmonic, relative sub- and ultra-harmonics, and relative 1/3, 2/3 sub- and ultra-harmonics, broadband noise indicative of inertial cavitation) become elevated with increasing pressure. Specifically, power in these peaks became statistically significant at 3 MPa. As a metric of IC dose, the integrated IC metric over all the bursts is shown in *Figure 6C*, showing that there is a significant difference in dose at 3 MPa. IC dose was also found to significantly increase from 1-3 and 2-3 MPa.

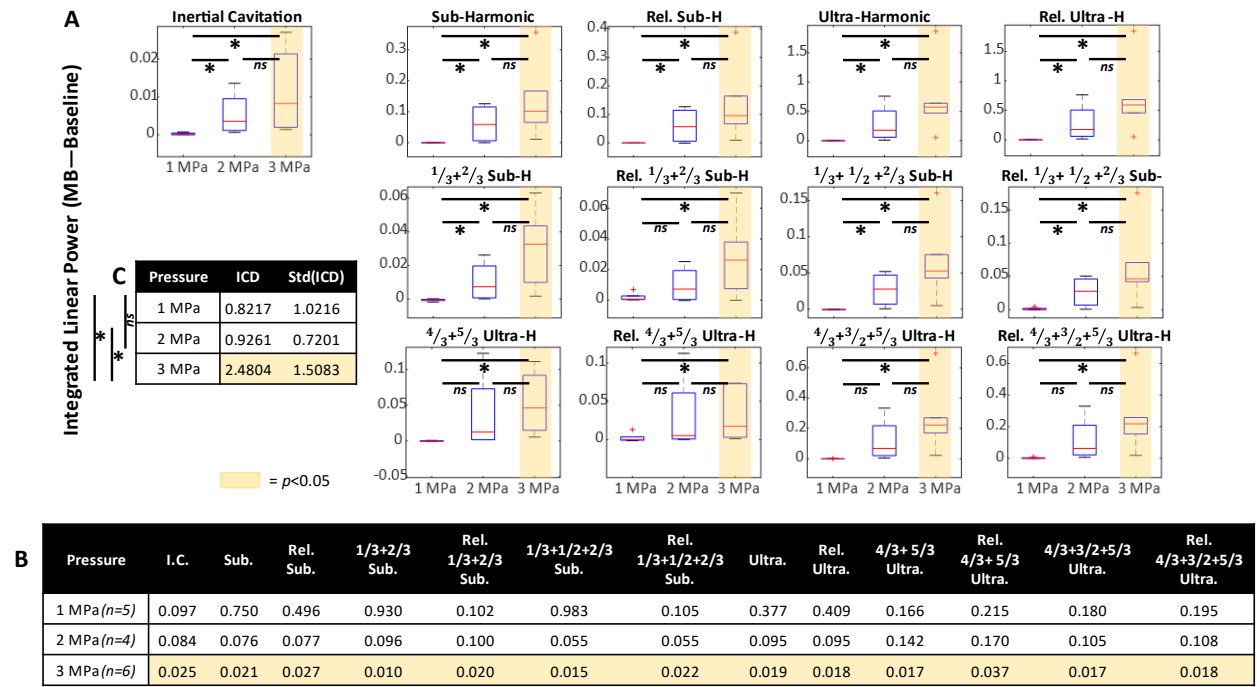

*Figure 6. Cavitation data summary.* (**A**) Box and whisker plots displaying the integrated power (of treatment with baseline subtracted) for each pressure group at various frequency peaks for the first burst in tumor tissue. (**B**) Table of the mean power ratios for pressure groups at each frequency peak of interest. The red crosshairs indicate outliers. Groups with statistically significant ($p<0.05$) elevation above baseline are highlighted in yellow. (**C**) IC dose (ICD) as a function of pressure. * $p<0.05$, n.s. = not significant.

In normal tissue, however, fewer frequency bands of interest reached statistical significance (*Figure S3*). At 1 MPa, power in the relative 1/3 and 2/3 subharmonic was significant. At 2 MPa, power in the 4/3 and 5/3 ultraharmonic bands became significant. At 3 MPa, only power in the 1/3 and 2/3 subharmonic and the relative 1/3 and 2/3 subharmonic peaks was significant.

Cavitation signal persistence was variable across treatment duration in both tumor-affected and normal tissue (***Figure S4***). Power in the frequency bands of interest generally decayed quickly over the first 3 bursts, then either continued to decay slowly (subharmonics) or remained plateaued (inertial cavitation, ultraharmonics) for the duration of the 18 burst treatment. Within the first burst, inertial cavitation was found to begin elevated and decline rapidly within the first ~100 µs of the 5 ms pulse length. The other harmonic peaks displayed variable behavior over the pulse length. In general, the subharmonics were found to increase over the first 1 ms of the pulse length, then decay, then again rise later (at ~3 ms) and fall again. The ultraharmonics were also found to generally rise over the first 1 ms and then fall, with only the third order ultraharmonics rising again closer to the end of the pulse length, at 4 ms.

**DISCUSSION**
A growing body of preclinical work suggests that AVUS has considerable potential to enhance the antitumor effects of conventional methods of chemotherapy [38, 15, 30], antiangiogenic therapy [39, 40], immunotherapy [41, 42], and radiotherapy [43, 44, 45, 46, 47, 48]. Initial clinical trials are underway investigating AVUS-mediated radiosensitization ( [72], NCT04431674, NCT04431648), albeit using systems that were not developed for cavitation-based FUS. As is the case for cavitation-based FUS drug delivery, to fully harness the potential of AVUS it is necessary to gain an understanding of the impact of the biological effects induced under different exposure conditions and their link with cavitation signatures. The present study was performed to elucidate FUS effects at the microvascular level in real-time in an acute setting, with the goal of providing information that informs the rational development of exposure and control systems for AVUS.

The most prominent immediate events observed during sonication (via XYT time acquisitions) were focal microvascular disruptions as indicated by localized dye leakage, which increased in prevalence with pressure and became more spatially diffuse over time. The extravasation of dye or nanoparticles from microvessels during FUS has been previously observed with intravital microscopy in both normal [73, 59, 74] and tumor [60, 53, 63, 69] tissue. These events accrued during the course of successive pulses following MB injection, with subsets resulting in flow cessation in the same vessel within seconds. Apparent 'vascular disruption' or rupture has been previously observed (***Table S1***), though few studies provide insight into possible flow cessation associated with these events. Although leakage from the first dye injection enabled insight into focal disruptions, its diffusion over time did not permit reliable quantification of sustained vascular shutdown, which was enabled by the volumetric analysis of the second dye distribution. Collectively these data indicated that only a subset of observed focal events were sufficiently strong to induce sustained vascular damage, with the remainder only resulting in enhanced permeabilization and transport of dye to the extravascular space. The proportion of focal events associated with vascular shutdown increased with pressure (79% at 3 MPa), and the variability of effects is nominally consistent with the polydisperse nature of Definity and the expected influence of vessel size on MB response. Conversely, only a fraction of vessels that were observed to have no flow (shutdown) also had a focal event within them (52% at 3 MPa). If it is assumed that a MB event that is sufficiently strong to induce vascular shutdown of the vessel would also produce dye leakage (to be classified as a focal disruption event), then this suggests that the flow cessation is not due to an event occurring within the vessel. Indeed it should be noted that 52% is an upper bound of the number of vessels that are shutdown due to a focal event within them, as it is possible that the focal disruption event within them may have simply been a permeabilization event. It is therefore hypothesized that flow cessation in vessels without focal disruption present is due to a microvascular network effect: it is associated with them being distal to or dependant on vessels that have been shutdown due to disruptive focal events within them. That is, this implies that it is not required that all vessels be directly 'shutdown' by an event within them due to the cascading effects of discrete events within particular vessels in the network.

Collectively considering all vessels, there was a progressive increase in vascular shutdown in tumors with pressure, being significantly higher at 3 MPa than 1 and 2 MPa (***Figure 4B***). It was also found that smaller vessels were preferentially affected: Vessels experiencing shutdown were strictly below 20 µm in diameter at 1 MPa, while at 3 MPa more vessels of larger diameters (up to 71 µm) were affected (***Figure 3C-D***). There is considerable microscopy-based evidence in normal vascular beds of localized microvascular damage *in vivo* due to FUS, in particular with demonstrations of capillary rupture (***Table S1***, [75, 76, 59, 77, 78]). In many circumstances, this work was carried out with a view to assessing potential bioeffects of MBs for conditions relevant to diagnostic MB contrast imaging with short imaging pulses. To date there is little direct quantitative evidence of flow cessation as a function of microvessel sizes, in normal or tumor tissue, under conditions that may be relevant to assessing AVUS therapy. In [51] the use of single short (3 cycle) pulses for acute vascular damage was assessed and observed that smaller vessels were preferentially affected and that the size increased with pressure (3-7 MPa). Notably in [51], at 3 MPa (our maximum and their minimum pressure) there were far fewer and smaller vessels shutdown than in the present study. While a direct comparison is confounded to an extent by differences in methodology, a likely factor in accounting for our increased level of vascular shutdown at lower pressures is the use of longer pulses, which were done here for compatibility with narrowband therapy array systems.

Vascular shutdown was found to occur more frequently in tumor-affected vessels, with greater severity as a function of increasing pressure. At 1 MPa, 38.5% of vessels exhibited sustained shutdown in tumors compared to 0% in normal tissue; while at 3 MPa, 97.1% of affected tumor vessels experienced sustained shutdown compared to 75.7% in normal vasculature. This effect is likely due to the abnormal morphology exhibited by tumor vessels, as angiogenic tumor vessels are typically immature. This may have implications for reducing off-target effects for therapy, though the tissue type in question needs to be considered (for example, the vulnerability of brain tissue as indicated by the varying parameters used, outlined in *Table S1*).

This study used FUS exposures at ~1 MHz, with a pulse length of 5 ms, and pulse repetition period of 10 s for a 3 min duration, at 1, 2, or 3 MPa. These parameters fall in the range of protocols used in other AVUS studies that have been carried out in bulk tumor or normal tissue (frequency 0.24-5 MHz; modes of continuous or pulsed waves of 0.1-100 ms with pulse repetition frequencies of 1-10 kHz; peak negative pressures of 0.18-5 MPa; *Table S1*). A direct comparison of the results obtained here with previous work is complicated by factors such as MB dose, FUS pressures, and the metrics and timepoints of assessing antivascular effects. In terms of dosing, the levels used here are higher than those employed clinically in a diagnostic ultrasound setting but are within range of what is being utilized in initial clinical trials and in our prior preclinical work. Notably in our previous studies [41, 15, 30], a peak negative pressure on the order of 1.6 MPa was applied to induce vascular shutdown, though due to the use of a loosely focused transducer to enable complete coverage of subcutaneous tumors (f-number 4) non-linear propagation resulted in peak-positive pressures on the order of 3.5 MPa. Antivascular effects have been reported in preclinical tumors at lower pressures but with substantially higher MB doses (*Table S1*). It is also notable that our metric here is sustained vascular shutdown at the 1 h point, whereas many other studies assess vascular shutdown and tissue death (necrosis, apoptosis) at longer timescales (24 h). However, it has been shown that vascular shutdown immediately following treatment of tumor tissue has been associated with subsequent tissue necrosis [30, 34].

This study also permitted a comparison of observed microvascular effects with cavitation data derived from the microvascular bed. As a general point, at higher pressures inertial cavitation was observed alongside distinct peaks at subharmonic and ultraharmonic frequencies. The latter included not only the widely reported order 1/2 subharmonic and 3/2 ultraharmonic but also more complex order 1/3, 2/3, 4/3 and 5/3 harmonics. These can in principle arise from spherically symmetric bubble oscillations [79] and have also been reported in the context of laser induced bubble clouds [80]. In a FUS setting, these have previously been observed anecdotally in the spectra in subcutaneous mouse tumors [30], and recent evidence from a phantom study suggests that they only arise from bubbles in microvessels less than 50 µm in diameter at higher pressures (2 and 3 MPa) [81]. Inertial cavitation was found to increase significantly as a function of pressure and this was accompanied by increases in vascular shutdown. Previous work in mm-sized vessels has linked vascular damage with inertial cavitation [11]. This is also consistent with a wide range of studies that link the *in vivo* presence of inertial cavitation (detected in larger sample volumes and focal regions) with vascular bioeffects such as erythrocyte extravasation. Recently in [67], inertial cavitation levels were linked to subsequent non-thermal necrosis regions induced by vascular shutdown. In [53] microvascular damage as measured by microscopy was linked to inertial cavitation as measured in a phantom under similar conditions. Aside from inertial cavitation, we have shown significantly elevated levels of a variety of sub- and ultra-harmonics, alone or in combinations, at increasing pressure levels. Notably, this is also the case for the 'relative' peaks, which are measured relative to the inertial cavitation level to draw conclusions about the peaks. It also warrants comment that the lack of significance of the inertial cavitation, subharmonics and ultraharmonics relative to the baseline until 3 MPa appears to depart from previous FUS drug delivery and MB characterization studies in terms of the pressure onset, which may be due to several factors. One is that the expected signal strength in this study is lower due to the smaller volume of tumor present in the beam (compared to the entire therapy focus being filled with tumor tissue). A second is that we are examining a microvascular bed (no contributions from larger vessels as is the case with larger tissue volumes) and microvessels are expected to have a higher pressure threshold for the onset of inertial cavitation [82] and sub- and ultra-harmonics [81]. The onset of significance of the different order (1/3 and 2/3) subharmonics and (4/3 and 5/3) ultraharmonics at a pressure level that corresponds to high levels of vascular disruption suggest that their use for control and dose monitoring warrants investigation.

The current study also indicated a high level of variability in terms of cavitation signal persistence: At 2 MPa with just under 25% shutdown, inertial cavitation only persisted for ~0.1 ms and for 1-2 bursts, while sub- and ultra-harmonic peaks (and their associated 1/3, 2/3 peaks, when they occurred) generally persisted over the full 5 ms pulse length, but were short-lived for only 1-2 bursts of the 18 burst (3 min) total duration. At 3 MPa, inertial cavitation again persisted for ~0.1 ms during bursts but persisted (with some decay) across the full 18 burst treatment duration. Other harmonic peaks again persisted over the full 5 ms pulse length, for variable repeated bursts ranging from 2-18 of the total 18 burst duration. This indicates that while 5 ms pulse lengths are sufficient for vascular shutdown, shorter pulse lengths of 0.1 ms are worth exploring. It additionally indicates that a 3 min total treatment duration is not strictly necessary to achieve vascular shutdown, and shorter treatment timelines may suffice. The recorded cavitation signal patterns noted here may aid in establishing a more specific AVUS treatment protocol, and further indicate a potential basis for acoustic feedback control to monitor and maximize treatment efficacy.

The observed vessel events and their time-course following AVUS shed some light on underlying vascular disruption mechanisms. As noted above, an insight provided here is that vascular shutdown appears due to a combination of vessels that have had a bubble event occur within them that induces shutdown, as well as vessels where flow has ceased due to shutdown elsewhere in the network. The following discussion relates to the former case. One proposed model in the literature is that AVUS damages the endothelium, inducing platelet aggregation and vessel thrombosis [60, 32], which is derived from data using endothelial cell targeted MBs and shorter pulses at higher pressure. However, clot formation and vascular occlusions were rarely visualized here, occurring only 6.8% of the time at 3 MPa. This may be due to may be due to different exposure conditions (shorter, higher pressure pulses were employed in [32]) or due to the use of endothelial cell targeted MBs in [32]. Thrombus, alongside endothelial cell damage was also observed in [11], although this was in mm-sized vessels. This study also visualized numerous cases where blood flow shutdown was transient with no obvious structural changes, pointing perhaps to mechanisms of vasospasm or vasoconstriction. These effects too, however, were only visualized in up to ~5% of vessels at higher pressures. Another proposed model in the case of combined AVUS and radiotherapy is that endothelium perturbation can upregulate ASMase and ceramide [47, 64], leading to endothelial cell death. Our results do not exclude that this is happening, though it should be noted that the timescale of shutdown in the present study (seconds) is far shorter than the timescale of apoptosis induction and as such this is unlikely to be the cause of shutdown here. As the radiosensitization work was conducted at lower pressures (500-750 kPa) and with higher MB doses, it is reasonable to consider that different mechanisms can be responsible for vascular disruption, depending on the exposure conditions. Indeed the XYT data indicate that events that give rise to vessel shutdown are localized and involve the ejection of dye into the extravascular space which is consistent with previous reports of rupture in capillaries [75, 76, 59, 78] and in tumor vessels [53]. In [75] evidence was provided of endothelial cell injury near rupture sites that had erythrocyte extravasation. *Ex vivo* experiments with high-speed camera measurements also suggest that capillary walls and endothelial cells can be damaged with violent bubble oscillations [77, 83].

The XYT data also demonstrate that focal disruption is also frequently accompanied by substantial tissue deformation. While the frame rate (8 Hz) is orders of magnitude too low to visualize tissue deformation occurring at the timescale of bubble oscillation, it is evident that this motion and deformation extends into and well beyond the perivascular space and can take a number of image frames to recover to its original position. The implications of this degree and type of motion and strain on vascular effects and possible fluid transport [84] warrant investigation. Finally, as AVUS has been shown to be particularly effective when combined with other therapeutic agents [38, 15, 30, 41, 64]; the enhanced extravasation effects into the tumor extravascular space and greater bioavailability promoted by FUS at higher pressures may facilitate such synergistic outcomes.

Study limitations should be noted: While studies were conducted over the same timeframes, tumor growth varied across mice in terms of tumor size, vessel involvement, and location within the dorsal window. As imaging and sonication were physically constrained over a region of interest in the centre of the window chamber, such differences in tumor size and location may have affected the morphology and strength of the visualized vessels. The vessel size distributions were also heterogeneous, making interpretation of cavitation results and their correlation with vascular events difficult. It is also noteworthy that pilot experiments determined that 3 MPa sonications in the absence of MBs were capable of inducing a small degree of vascular disruption, presumably due to endogenous cavitation seeds. For this reason, baseline sonications were limited to a shorter duration of 2-3 bursts, which did not result in notable disruption. It should also be noted that several ring transducers were fabricated to expedite experiments due to their fragile nature; while these were each calibrated prior to experiments, minor pressure distribution differences existed between different rings which may have contributed to variability within pressure groups.

**CONCLUSION**
AVUS has experienced early preclinical success however, the establishment of standard protocols and control approaches has been impeded by fundamental questions about exposures, vessel types and sizes involved, and the nature of acoustic cavitation resulting in vascular damage. Here, vessel type (normal and tumor-affected), caliber, events, and viability were assessed during ultrasound exposure (1, 2, and 3 MPa) of circulating exogenous MBs, and associated cavitation was monitored. Vascular events were found to occur with greater incidence in smaller vessels, with more severity as a function of increasing pressure, and with higher incidence in tumor-affected vessels. Vascular shutdown was observed to be due to a combination of focal disruption events and network related flow inhibition. Detected acoustic emissions indicated elevated broadband emissions along with sub- and ultra-harmonics and their associated 1/3 and 2/3 peaks with increasing pressure. Identification of broadband emissions and distinct sub- and ultra-harmonics taken collectively with identified vascular events provide an improved mechanistic understanding of AVUS at the microscale, with implications of establishing a specific treatment protocol and control platform.

**ACKNOWLEDGEMENTS**
The authors thank S. Bulner for performing tumor cell culture; V. Chan and S. Rideout-Gros for advising on the dorsal skinfold window chamber implantation surgeries; M. A. O'Reilly for designing the PVDF receive transducer; A. Wright for constructing the ring PZT transducers and PVDF receive transducers; M. A. Santos for pilot study guidance; and A. Dorr for her support in using the multiphoton microscope. Funding was provided by the Canadian Institutes for Health Research, Ontario Research Fund, and the Provost's PhD Engagement Fund Sunnybrook Award.